\documentclass[a4paper,12pt]{article}           
\usepackage{ucs}
\usepackage[utf8x]{inputenc}
\usepackage[T2A]{fontenc}
\usepackage{amssymb,amsfonts,amsmath,mathtext,amsthm}
\usepackage{graphics}
\usepackage{graphicx}

\DeclareMathOperator{\arccot}{arccot}

\begin{document}

\title{\large{\bf{ Representation of the gravitational potential of a 
level ellipsoid by a simple layer}}}

\author{\normalsize{D.~V. Milanov\thanks{e-mail: danila.milanov@gmail.com}}}
\date{\normalsize{St. Petersburg State University, St. Petersburg, 198504 Russia}}

\maketitle

\begin{abstract}
A closed-form expression is obtained for the density of a simple layer, equipotential 
to an oblate level ellipsoid of revolution in an outer space. The potential of any 
level spheroid of positive mass with the inward direction of attracting force on its surface
can be represented in this way. A family of density functions defined on the whole
volume of a level ellipsoid of revolution is found. Several density examples are considered.
\end{abstract}

\section{Introduction}\label{intro}
The necessary condition of the equilibrium of a planet is the constancy of potential of attracting
force on the planet's surface:
$$
W(x, y, z) = V(x, y, z) + \frac{\omega^2}{2}\left(x^2 + y^2\right) = const
$$
or
\begin{equation}\label{equilibrium}
V(x, y, z) = V_0 - \frac{\omega^2}{2}\left(x^2 + y^2\right),
\end{equation}
 where $V(x, y, z)$ is the gravitational potential at a point with Cartesian coordinates $(x, y, z)$, 
 $\omega$ is the angular velocity of rotation of the planet about the $z$ axis, 
 $V_0$ is the value of $V$ at the pole.
The oblate ellipsoid of revolution (spheroid) 
\begin{equation}\label{ellipsoid-cart}
\frac{x^2 + y^2}{a^2} + \frac{z^2}{c^2} = 1
\end{equation}
with semi-major and semi-minor axes $a > c$, 
serves as a relatively simple and at the same time close to reality model of a planet's form
\cite{Hofmann-Wellenhof-2006}.

The outer gravitational potential of spheroid \eqref{ellipsoid-cart}, 
satisfying the boundary condition \eqref{equilibrium} (the level spheroid), 
is uniquely determined by
$V_0$ and $\omega$ or another pair of real constants corresponding to those.
Due to linearity of the Laplace operator, it is convenient to choose the body mass $M$
(the limit of $rV$ at infinity) as the first constant. The potential then non-trivially depends 
only on the second one which is proportional to $\omega^2/M$.

\cite{Pizzetti-1913} obtained the explicit expression for the outer potential of a level spheroid. 
This expression is given in the next section \eqref{poten-explicit}.
The coefficients of expansion of this potential into the Laplace series 
were found in \cite{Kholshevnikov-2018}.

In this article the representation of the outer potential of a level spheroid in 
the form of a potential of simple layer is built. In the section \ref{simple-layer}
the explicit expression is given for the family of density functions defined on the 
surface of spheroid, whose potential solves the Dirichlet problem with the boundary
condition \eqref{equilibrium}. The family is parameterized by a real number. 
It is shown in the section \ref{a-b-ranges}, that the family contains density
of an equipotential simple layer for any level ellipsoid of positive mass with
 inward direction of attracting force on its surface. As opposed to a volume density,
the continuous density of a simple layer is uniquely determined by its outer potential.

Despite the fact that the general form of the potential of level ellipsoid is known,
it is unclear for which values of $M$ and $\omega$ there exists a mass distribution
in ellipsoid that induces the potential with given parameters. Examples of level spheroids
known to us are exhausted by a homogeneous one, an ellipsoid with confocal mass stratification
\cite{Kondratiev-2003}, and a sum of one of those types with a body of zero attraction \cite{Pizzetti-1913}.
In all cases above the outer potential coincides with the potential of a homogeneous ellipsoid,
providing a single value of the parameter $\omega^2/M$.

The family of densities defined on the whole volume of a level ellipsoid is constructed
in the section \ref{volume-density}. The family is parameterized by three functions.
Those mass distributions do not solve completely the problem of existence of a density
for a given potential, but provide a material for estimates of the parameter range.
In particular, the family contains density for which the value of $\omega^2/M$ is 
separated from zero for an arbitrary small eccentricity. This one and other examples are
given in the section \ref{examples}.


\section{The outer potential of the level spheroid}
The object of our study is the ellipsoid of revolution 
\eqref{ellipsoid-cart} with various mass stratifications.
Let us introduce some notation which describe the shape of the ellipsoid
$$
\varepsilon^2 = 1 - \frac{c^2}{a^2}, \quad \chi = \frac{\sqrt{1-\varepsilon^2}}{\varepsilon}.
$$
Instead of the angular velocity of rotation $\omega$ we use the dimensionless Clairaut parameter:
\begin{equation}\label{clairaut}
q = \frac{3\omega^2}{4\pi \mathcal G\overline{\varrho}} = \frac{\omega^2a^2c}{\mathcal GM},
\end{equation}
where $\mathcal G$,  $M$, and $\overline{\varrho}$ are the gravitational constant, the mass of the body and its mean density.

Further we mainly work with spheroidal coordinates $u, v, \lambda$, related to Cartesian
ones with the following formulae \cite{Korn-1968}:
\begin{equation}\label{elliptic}
 \begin{split}
  x& =\mathfrak b\sqrt{1+u^2}\sqrt{1-v^2}\cos\lambda,\\
  y& =\mathfrak b\sqrt{1+u^2}\sqrt{1-v^2}\sin\lambda,\\
  z& =\mathfrak b u  v .
 \end{split}
\end{equation}
The azimuth  $\lambda$ varies from $0$ to $2\pi$. The coordinate surface $\lambda = const$
is a half-plane  containing the applicate axis. The surface $v = const$ is a one-sheeted hyperboloid
of revolution; $-1\leqslant v \leqslant 1$. The third coordinate $u\geqslant 0$ defines an
oblate spheroid with its focal distance equal to $\mathfrak b$.

The equation of ellipsoid \eqref{ellipsoid-cart} takes the form $u = \chi$ in 
spheroidal coordinates with the parameter $\mathfrak b = a\varepsilon$. The element of
surface of the ellipsoid and the element of its volume are expressed as following:
\begin{equation}\label{elliptic-SV}
dS = a^2\sqrt{1-\varepsilon^2 + \varepsilon^2v^2}dv d\lambda,\quad
dV = \mathfrak b^3(u^2 + v^2)du dv d\lambda.
\end{equation}

The outer gravitational potential of spheroid $u=\chi$, satisfying the boundary condition 
\eqref{equilibrium}, is uniquely determined by constants $M$ and $\omega$ and can be
expressed in elementary functions of coordinates. Following the notation from \cite{Kholshevnikov-2018},
\begin{equation}\label{poten-explicit}
V(u, v) = \frac{\mathcal GM} {a}\left(\frac{1}{\varepsilon}G_1(u) + A_0G_2(u)P_2(v)\right),
\end{equation}
where
\begin{equation}\label{poten-explicit-q}
\begin{split}
   G_1(u) = \arccot u,\quad G_2(u) &= \left(3u^2 + 1\right)G_1(u) - 3u, \quad A_0 = \frac{q}{3G_3(\varepsilon)}, \\
   G_3(\varepsilon) &= G_2(\chi)\sqrt{1-\varepsilon^2},
\end{split}
\end{equation}
$P_2(v) = (3v^2 - 1)/2$ is the Legendre polynomial of the second order.

The outer potential \eqref{poten-explicit} can be expanded into the Laplace series:
\begin{equation}\label{laplace}
V(x, y, z) = \frac{\mathcal GM}{a}\sum\limits_{n=0}^{\infty}I_{2n}\frac{a^{2n+1}}{r^{2n+1}}P_{2n}\left(\frac{z}{r}\right),
\end{equation}
where $P_{2n}$ is the Legendre polynomial of order $2n$, $r = \sqrt{x^2+y^2+z^2}$, 
$I_{2n}$ are numeric coefficients (Stokes coefficients). 
Given the density $\varrho$ of the ellipsoid with support $E$ (either a volume or a surface),
Stokes coefficients can be expressed with the following formula: 
$$
I_{2n} = \frac{1}{Ma^{2n}}\int\limits_{E} \varrho r^{2n} P_{2n}\left(\frac{z}{r}\right)dV.
$$
The coefficients $I_{2n}$ of the Laplace series for the level spheroid were calculated in \cite{Kholshevnikov-2018}:
\begin{equation}\label{i2n}
I_{2n} = (-1)^{n}\frac{2An + 3}{(2n+1)(2n+3)}\varepsilon^{2n}, \quad A = 1-2A_0\varepsilon.
\end{equation}
Based on the last result we deduce one more representation of the potential \eqref{poten-explicit}
in the next section.
\section{Representation by the potential of the simple layer}\label{simple-layer}
Let $E$ be the ellipsoid of revolution defined by the equation $u=\chi$ in spheroidal coordinates
\eqref{elliptic}.
Let us define the surface density on $E$ as the function
\begin{equation}\label{sl-rho}
\varrho(v, \lambda) = \frac{1+Bv^2}{\sqrt{1 - \varepsilon^2 + \varepsilon^2v^2}},
\end{equation}
where $B \geqslant -1$ is a constant.

Further we show that Stokes coefficients of potential of $E$ with density \eqref{sl-rho}
take the form \eqref{i2n} with the parameter $A$ depending on $B$ and $\varepsilon$. Thus we
prove the coincidence of the outer potential of $E$ with the function \eqref{poten-explicit}.

Let us calculate the mass of the ellipsoid
\begin{equation}\label{sl-mass}
M = \int\limits_{E}\varrho(v, \lambda)\,dS = 
a^2\int\limits_{0}^{2\pi}d\lambda \int\limits_{-1}^{1} (1 + Bv^2)\,dv = 
\frac43\pi a^2(B+3).
\end{equation}
Stokes coefficients of spheroid's outer gravitational potential are
$$
I_{2n} = \frac{1}{Ma^{2n}}\int\limits_{E}\varrho r^{2n} P_{2n}\left(\frac{z}{r}\right)\,dS,
$$
where, at the surface $u=\chi$,
$$
r = a \sqrt{1 - \varepsilon^2v^2},\quad
\frac{z}{r} = \frac{v\sqrt{1-\varepsilon^2}}{\sqrt{1 - \varepsilon^2v^2}}.
$$
Taking into account the symmetry in $\lambda$,
\begin{equation}\label{I2n-int-theta}
I_{2n} = \frac{3}{2(B+3)}
\int\limits_{-1}^{1}\left(1+Bv^2\right)\left(1-\varepsilon^2v^2\right)^n 
P_{2n}\left(\frac{v\sqrt{1-\varepsilon^2}}{\sqrt{1 - \varepsilon^2v^2}}\right)dv.
\end{equation}
The change of variable
\begin{equation*}
\begin{split}
 t=\frac{v\sqrt{1-\varepsilon^2}}{\sqrt{1-\varepsilon^2 v^2}},\quad
  v&=\frac{t}{\sqrt{1-\varepsilon^2+\varepsilon^2t^2}},\quad
   1-\varepsilon^2v^2=\frac{1-\varepsilon^2}{1-\varepsilon^2 + \varepsilon^2t^2},\\
   dt&=\frac{\sqrt{1-\varepsilon^2}dv}{(1-\varepsilon^2v^2)^{3/2}}
\end{split}
\end{equation*}
brings \eqref{I2n-int-theta} to a form
\begin{equation*}
\begin{split}
I_{2n} &= \frac{3(1-\varepsilon^2)^{n+1}}{2(B+3)} \int\limits_{-1}^1 
\frac{1 - \varepsilon^2 + (B+\varepsilon^2)t^2}{(1-\varepsilon^2 + \varepsilon^2t^2)^{n+5/2}}P_{2n}(t) dt
\\&=\frac{3}{2(B+3)\sqrt{1-\varepsilon^2}}\, F\left(\frac{B+\varepsilon^2}{1-\varepsilon^2}, \frac{\varepsilon^2}{1-\varepsilon^2}, n\right),
\end{split}
\end{equation*}
where
\begin{equation}\label{F-5-2}
F(x, y, n) = \int\limits_{-1}^1\frac{1+xt^2}{(1+yt^2)^{n+5/2}}P_{2n}(t) dt.
\end{equation}
The last integral was calculated in \cite{Kholshevnikov-2017}:
\begin{equation*}
F(x, y, n) =\frac{2(-1)^{n}y^{n}}{(2n+1)(2n+3) (1+y)^{n+3/2}}
\left(2n\left(2+y - \frac{x}{y}\right) + 3 +2y+x\right).
\end{equation*}
After substitution and subsequent simplification we get
\begin{equation}\label{sl-i2n}
I_{2n} = \frac{(-1)^n\varepsilon^{2n}}{(2n+1)(2n+3)}\left(2An + 3\right),
\end{equation}
where
\begin{equation}\label{sl-A}
A = 3\frac{1-B\chi^2}{B+3},
\end{equation}
Coefficient \eqref{sl-i2n} coincides with \eqref{i2n}. Hence, the potential of the simple layer with density \eqref{sl-rho}
coincides with function \eqref{poten-explicit} with parameter $A_0 = (1-A)/2\varepsilon$ in the outer space. 
That is, the ellipsoid with the density \eqref{sl-rho} is the level one.


\section{Range of values of $B$}\label{a-b-ranges}
Let $A \neq -3\chi^2$ be a real number. 
The potential of simple layer with density \eqref{sl-rho} with parameter
$$
B = 3\,\frac{1-A}{A + 3\chi^2}
$$ 
coincides with the function \eqref{laplace} and consequently with the function \eqref{poten-explicit} with parameter
$$
    A_0 = \frac{1-A}{2\varepsilon} \neq \frac{3-2\varepsilon^2}{\varepsilon^3}.
$$
That is, the simple layer \eqref{sl-rho} represents potential of any level spheroid except 
the potential \eqref{laplace} with $A = -3\chi^2$. It can be shown \cite{Antonov-1988}, that the density of a simple layer
is uniquely determined by the outer potential in the class of continuous functions.

Let us consider the question about the special value $A = -3\chi^2$. 
Spheroids studied in the problems related with shapes of celestial bodies are those ones which have positive mass
and an inward direction of the attracting force on their surfaces.
The necessary condition of the last property was proposed in \cite{Kholshevnikov-2018} in a form of 
inequality for the Clairaut parameter \eqref{clairaut}:
\begin{equation}\label{q-range}
0 < q \leqslant G_6(\varepsilon),
\end{equation}
where
$$
G_6(\varepsilon) = \frac{3\varepsilon^2G_3(\varepsilon)}{2G_5(\varepsilon)},\quad
G_5(\varepsilon) = 3(1-\varepsilon^2)G_4(\varepsilon) - 3\varepsilon + 4\varepsilon^3,\quad
G_4(\varepsilon) = \sqrt{1 - \varepsilon^2}\arcsin \varepsilon.
$$
Let us express $q$ through $B$ using \eqref{sl-A}, \eqref{i2n}, and \eqref{poten-explicit-q}:
\begin{equation}\label{sl-q}
q = 3G_3A_0 = \frac{3G_3(\varepsilon)(1+3\chi^2)}{2\varepsilon}\frac{B}{B+3} = 
\frac{3G_3(\varepsilon)(3 - 2\varepsilon^2)}{2\varepsilon^3}\frac{B}{B+3}.
\end{equation}
Due to \eqref{sl-mass} the potential is positive at infinity only if $B > -3$. After solving the inequality \eqref{q-range} 
with respect to  $B$ we obtain:
$$
0 < B \leqslant \frac{3}{H(\varepsilon) - 1},
$$
where
\begin{equation}\label{sl-H}
H(\varepsilon) = \frac{G_5(\varepsilon)(3-2\varepsilon^2)}{\varepsilon^5} = 
\frac95 + 18\varepsilon^2\sum\limits_{k = 0}^{\infty} \frac{(2k)!!(k-5)}{(2k+7)!!}\varepsilon^{2k}.
\end{equation}
The series \eqref{sl-H} can be deduced from the expansion for $G_5$ given in \cite{Kholshevnikov-2018}.
It is shown in the appendix \ref{appendix} that $H(\varepsilon)$ is decreasing. Consequently,
\begin{equation}\label{sl-H-range}
1 = H(1)\leqslant H(\varepsilon)\leqslant H(0) = \frac95
\end{equation}
For the parameter $A$ of the series \eqref{laplace} the following estimate holds
\begin{equation*}
1 - \frac{1+3\chi^2}{H(\varepsilon)} \leqslant A < 1.
\end{equation*}
Taking into account \eqref{sl-H-range} and the condition $\varepsilon < 1$, one can conclude that $A > -3\chi^2$ is true.
That is, the  simple layer \eqref{sl-rho} represents potential of any level spheroid of positive mass
with the inward direction of attracting force on its surface.


\section{The level ellipsoid with volume density}\label{volume-density}
A slight modification of the formula \eqref{sl-rho} of the surface density and of consequent calculations
gives us a family of density functions defined on the whole volume of a level ellipsoid.

Let $E$ as before be the ellipsoid of revolution defined by equation $u=\chi$.
Let us define the volume density on $E$ as the function
\begin{equation}\label{rho-solid}
\varrho(u, v, \lambda) = 
f + g\frac{1+h v^2 }{u^2 + v^2},
\end{equation}
where $f, g, h$ are Riemann integrable functions of $u$.

The mass of ellipsoid is
$$
M = \int\limits_{E}\varrho d\tau = 
2\pi\mathfrak b^3\int\limits_{0}^{\chi}du \int\limits_{-1}^{1}
(u^2 + v^2)f + (1+hv^2)gdv = 
\frac43\pi\mathfrak b^3\int\limits_{0}^{\chi}
(3u^2 + 1)f + (3+h)gdu.
$$
Let us calculate Stokes coefficients of the outer gravitational potential of $E$.
\begin{equation*}
I_{2n} = \frac{1}{Ma^{2n}}\int\limits_{E}\varrho r^{2n} P_{2n}\left(\frac{z}{r}\right)\,dV
\end{equation*}
In spheroidal coordinates \eqref{elliptic}
$$
r^2 = \mathfrak b^2\left(1+u^2 -  v^2 \right),\quad
\frac{z}{r} = \frac{ u  v }{\sqrt{1+u^2 -  v^2 }}.
$$
This implies
\begin{equation*}
I_{2n} = \frac{2\pi \mathfrak b^{2n+3}}{Ma^{2n}}\int\limits_{0}^{\chi}du\int\limits_{-1}^{1}dv
\left(f + g\frac{1+h v^2 }{u^2 + v^2}\right)
\left(1+u^2 -  v^2 \right)^n 
 P_{2n}\left(\frac{ u  v }{\sqrt{1+u^2 -  v^2 }}\right)
\left(u^2 + v^2\right).
\end{equation*}
Let us change the variable in the integral in $v$.
$$
t =\frac{uv}{\sqrt{1+u^2 - v^2}},\quad dt = \frac{u(1+u^2)}{(1+u^2 - v^2)^{3/2}}dv,\quad
v^2 = \frac{t^2(1+u^2)}{u^2+t^2}
$$
$$
u^2+v^2 = \frac{u^4 + t^2(1+2u^2)}{u^2+t^2},\quad
1+u^2 - v^2 = \frac{u^2(1+u^2)}{u^2+t^2}.
$$
The integral of the term multiple of $f$, corresponding to the confocal mass stratification, gives
us, up to a factor, the Stokes coefficient of the homogeneous ellipsoid:
\begin{equation*}
\begin{split}
J_f &= f  \int\limits_{-1}^{1}
\left(u^2 + v^2\right)
\left(1+u^2 -  v^2 \right)^n 
 P_{2n}\left(\frac{ u  v }{\sqrt{1+u^2 -  v^2 }}\right)dv = \\
 u & (u^2(1+u^2))^{n+1/2} f 
 \int\limits_{-1}^{1}
 \frac{u^4+(2u^2+1)t^2}{(u^2+t^2)^{n+5/2}} P_{2n}(t)dt = 
u  (1+u^2)^{n+1/2} f \, F\left(\frac{2u^2 + 1}{u^4}, \frac{1}{u^2}, n\right),
\end{split}
\end{equation*}
where $F(x, y, n)$ is defined by equality \eqref{F-5-2}. 
After performing the transformations, we get
\begin{equation*}
J_f= \frac{2(-1)^nf}{(2n+1)(2n+3)} \frac{3u^4+4u^2+1}{u^2 + 1} = 
\frac{2(-1)^n}{(2n+1)(2n+3)}  \left(3u^2 + 1\right)f.
\end{equation*}
Calculations for the term multiple of $g$ give us
\begin{equation*}
\begin{split}
J_g &= g \int\limits_{-1}^{1}
\left(1+h v^2 \right)
\left(1+u^2 -  v^2 \right)^n 
 P_{2n}\left(\frac{ u  v }{\sqrt{1+u^2 -  v^2 }}\right)dv = \\
 u & (u^2(1+u^2))^{n+1/2} g 
 \int\limits_{-1}^{1}
 \frac{u^2+(1+(1+u^2)h)t^2}{(u^2+t^2)^{n+5/2}} P_{2n}(t)dt = \\
&\frac{(1+u^2)^{n+1/2}}{u} g \, F\left(\frac{1+(1+u^2)h}{u^2}, \frac{1}{u^2}, n\right) = 
\frac{2(-1)^n}{(2n+1)(2n+3)}  \left(2n(1-hu^2) + 3 + h\right)g.
\end{split}
\end{equation*}
Finally,
\begin{equation*}
I_{2n} = \frac{2\pi \mathfrak b^{2n+3}}{Ma^{2n}}\int\limits_{0}^{\chi}J_f + J_g du =
\frac{(-1)^n\varepsilon^{2n}}{(2n+1)(2n+3)}  \left(2An + 3\right),
\end{equation*}
where
\begin{equation}\label{solid-A}
A = \frac{4\pi\mathfrak b^3}{M}\int\limits_{0}^{\chi}\left(1 - hu^2\right)gdu = 
\frac{3\,\int\limits_{0}^{\chi}\left(1 - hu^2\right)gdu}
{\int\limits_{0}^{\chi}(3u^2 + 1)f + (3+h)gdu}\,. 
\end{equation}
Calculated coefficients coincide with the \eqref{i2n}. Consequently, the ellipsoid $E$
with the density \eqref{rho-solid} is the level spheroid.

\section{Examples}\label{examples}
\begin{figure}[!h]
  \includegraphics[width=\linewidth]{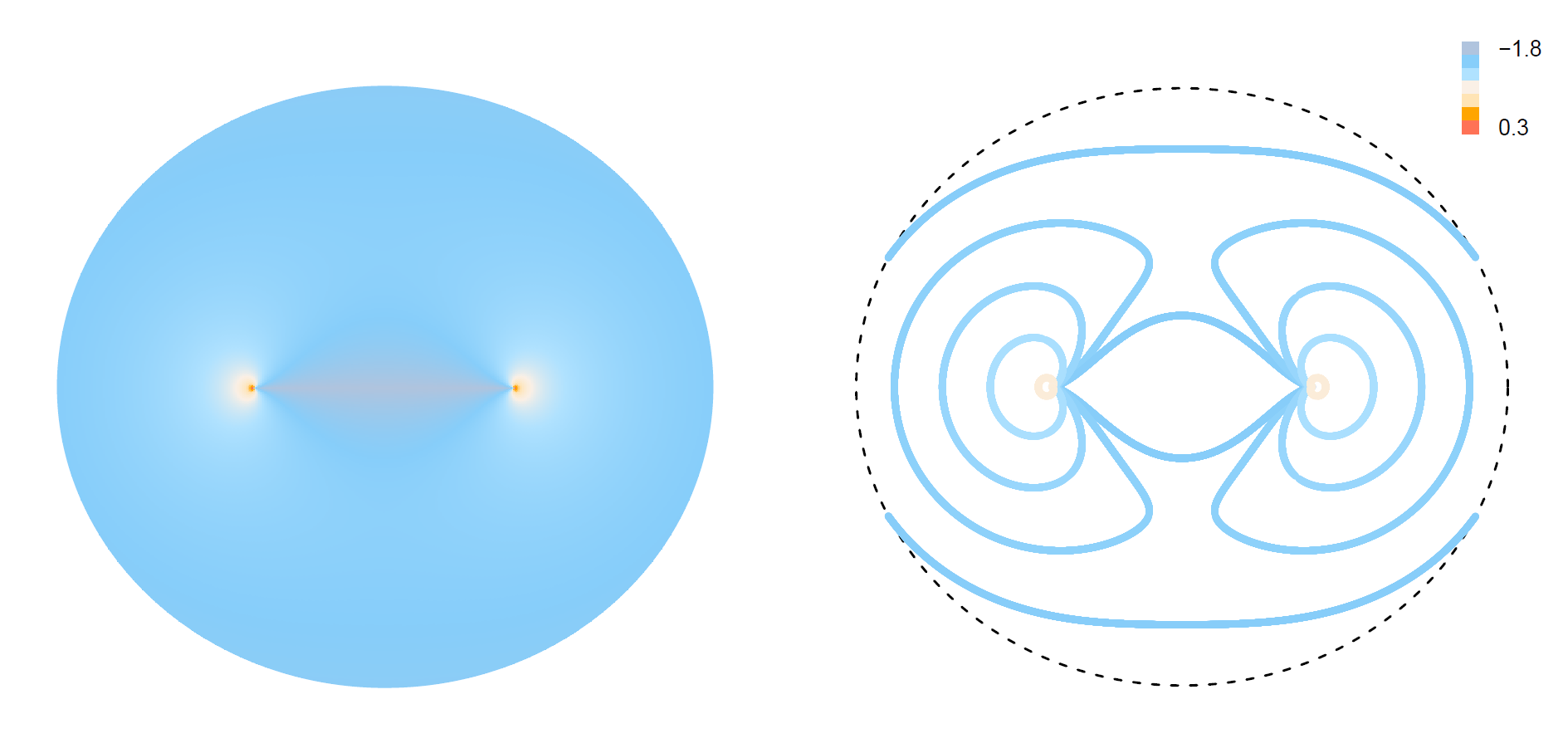}
  \caption{The section of the spheroid of unit mass with the eccentricity $\varepsilon=0.4$ and density
\eqref{rho-example-1}. On the left, all the points of the section are colored, on the right, only a few equidensites.
  The color of a point corresponds to the logarithm of the density.}
\label{fig:ex-1}
\end{figure}

{\bf 1.} Surfaces of the constant density (equidensites) in the ellipsoid with the mass stratification
\eqref{rho-solid} can have a variety of forms, depending on values of parameters $f$, $g$ and $h$.
In particular these surfaces are generally speaking not closed. For example, if $f(u)=h(u) = 0$ and $g(u)=u$ then
the density of ellipsoid and the equation of the equidensites are the following:
\begin{equation}\label{rho-example-1}
\varrho(u,v) = \frac{u}{u^2 + v^2}, \quad u^2 + v^2 = Cu,
\end{equation}
where $C$ is a constant. Obviously, infinitely many of level surfaces break off at the border of the figure.
The meridional section of the ellipsoid with eccentricity $\varepsilon=0.4$ and density \eqref{rho-example-1}
normalized so that the mass of ellipsoid is equal to one is shown on the figure \ref{fig:ex-1}.

The parameter $A$ of the Laplace series \eqref{laplace} of this ellipsoid calculated by formula 
\eqref{solid-A} equals to $1$. This is equivalent to $q=0$, that is the surface of ellipsoid with density
\eqref{rho-example-1} is the surface of the constant gravitational potential.
It easily follows from \eqref{solid-A}, that all ellipsoids with $f=h=0$ have the same property
independently from their eccentricity value.

\begin{figure}[!h]
  \includegraphics[width=\linewidth]{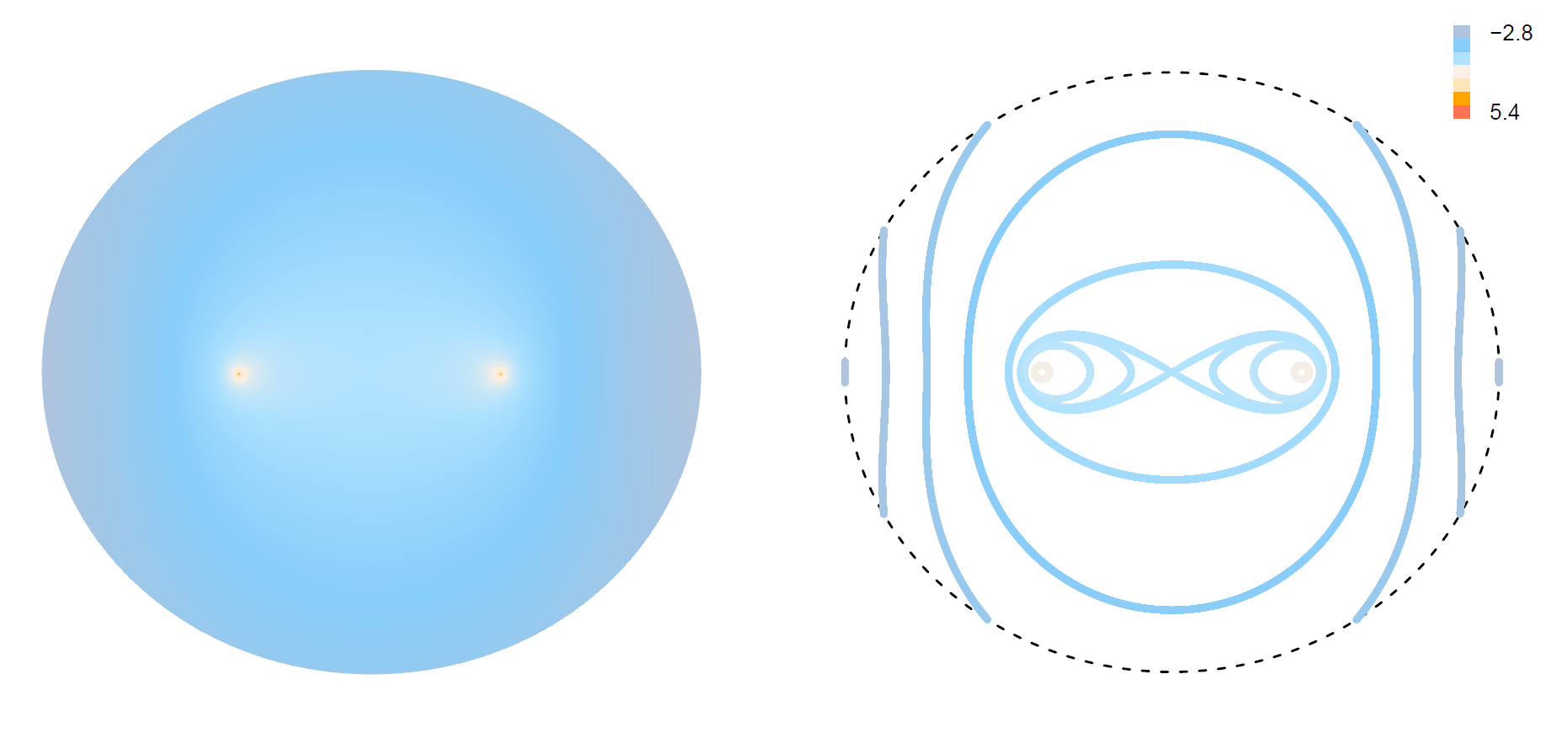}
  \caption{The section of the spheroid of unit mass with the eccentricity $\varepsilon=0.4$ and constant density parameters
$f=0$, $g = 1$, $h=2$. On the left, all the points of the section are colored, on the right, only a few equidensites.
  The color of a point corresponds to the logarithm of the density.}
\label{fig:ex-4}
\end{figure}

{\bf 2.} An interesting example of the ellipsoid with non-closed equidensites is given by
density parameters values $f=0$, $g = 1$, $h=h_0$, where $h_0$ is a positive constant. Parameters
$A$ and $q$ take the following form in this case:
$$
A = \frac{3-h_0\chi^2}{3+h_0},\quad q = \frac32\frac{h_0}{h_0+3}\frac{G_3(\varepsilon)}{\varepsilon^3}.
$$
With an infinitely small eccentricity, $A$ tends to $-\infty$, and $q$ tends to the value
$$
\lim\limits_{\varepsilon \to 0} q = \frac25\frac{h_0}{h_0+3} > 0.
$$
The section of the ellipsoid with parameter $h_0=2$ is shown on the figure \ref{fig:ex-4}.

\begin{figure}[!h]
  \includegraphics[width=\linewidth]{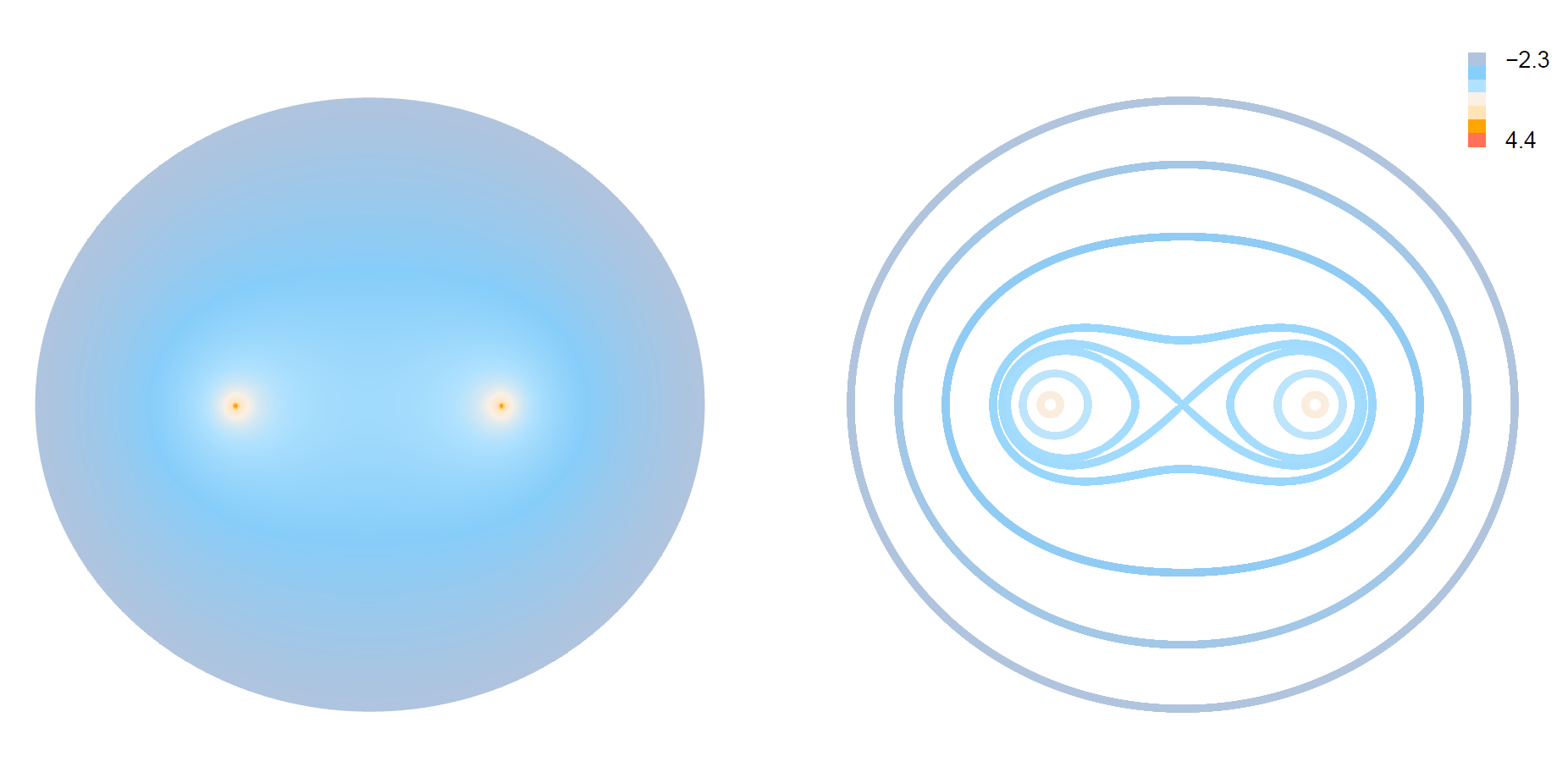}
  \caption{The section of the spheroid of unit mass with the eccentricity $\varepsilon=0.4$ and
  density parameters
$f = 0$, $g = \chi^{2}$, $h = \chi^{-2}$. On the left, all the points of the section are colored, on the right, only a few equidensites.
  The color of a point corresponds to the logarithm of the density.}
  \label{fig:ex-2}
\end{figure}

{\bf 3.} For equidensites to be closed, 
it is necessary that the density at the boundary be constant. This condition is equivalent
to one of the following two equalities:
\begin{equation}\label{closed-equidensite-condition}
g(\chi) = 0 \quad \text{or}\quad h(\chi) = \frac{1}{\chi^2},
\end{equation}
The section of the ellipsoid with $\varepsilon=0.4$ and density parameters
$f(u) = 0$, $g(u) = \chi^{2}$, $h(u) = \chi^{-2}$, satisfying the condition
\eqref{closed-equidensite-condition} is shown on the figure \ref{fig:ex-2}.

For this ellipsoid, the parameter  $A$ of the Laplace series \eqref{laplace} 
and the Clairaut parameter $q$ are the following:
$$
A = \frac{2\chi^2}{3\chi^2 +1}\approx 0.63, \quad q = \frac{\chi^2 + 1}{3\chi^2 + 1}\frac{G_3(\varepsilon)}{2\varepsilon}\approx 0.02
$$

\begin{figure}[!h]
  \includegraphics[width=\linewidth]{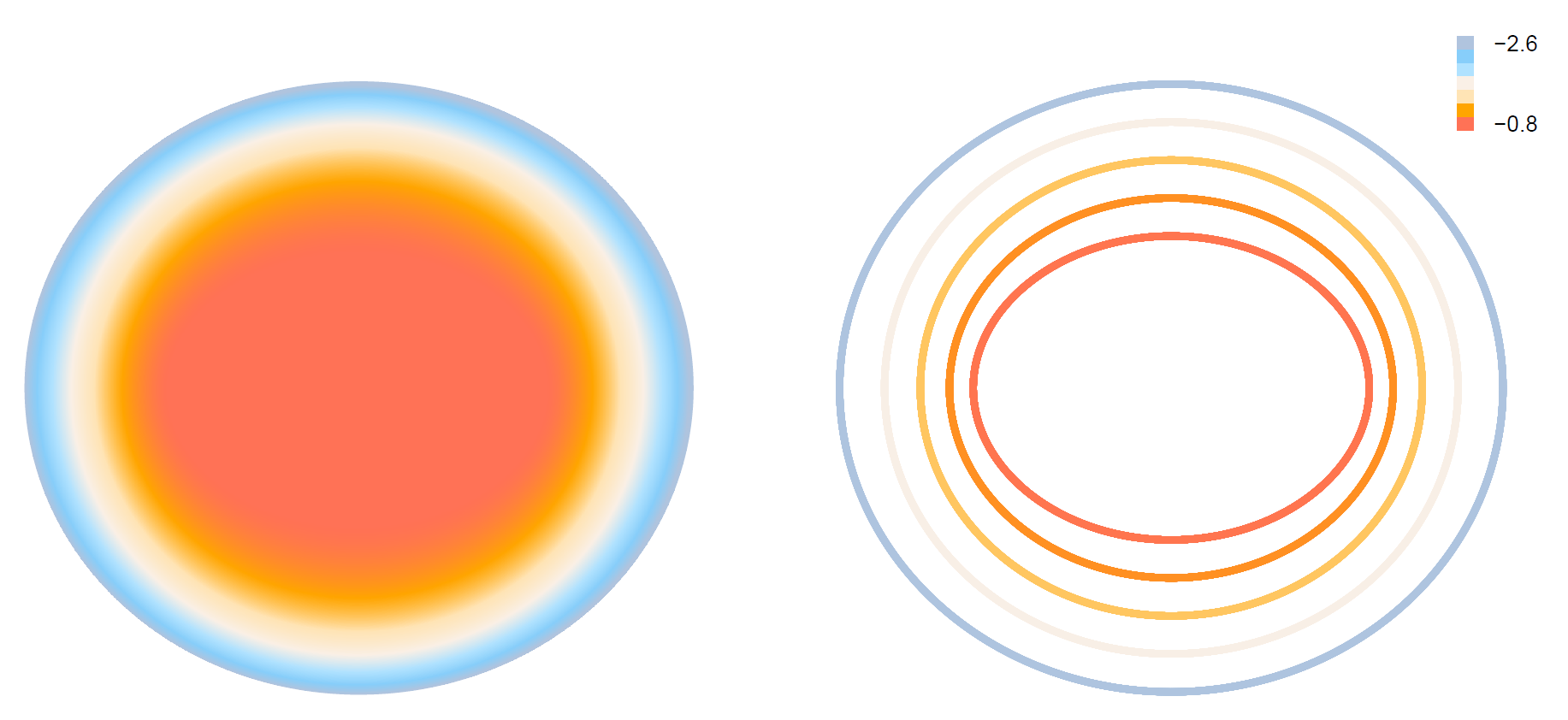}
  \caption{ The section of the spheroid of unit mass with the eccentricity $\varepsilon=0.4$ and density
\eqref{rho-example-3}. On the left, all the points of the section are colored, on the right, only a few equidensites.
  The color of a point corresponds to the logarithm of the density. The density inside the 
  ellipsoid $u \leqslant 1$ is constant.}
\label{fig:ex-3}
\end{figure}

{\bf 4.}  The function \eqref{rho-solid} has an essential singularity at the circle 
$u^2 + v^2 = 0$ of radius $\mathfrak b$, centered at the origin, 
located in the ellipsoid's equatorial plane. The only way to get rid of this singularity
is to put $g(u) = 0$ in a neighbourhood of the round $u=0$. In the example below,
the coefficients are selected so that the \eqref{closed-equidensite-condition} holds, and
the density is a continuous function together with its first derivatives.
\begin{equation}\label{rho-example-3}
 \begin{split}
f(u) = 1, \quad g(u) &= h(u) = 0,\, \text{if} \, u \leqslant 1,\\
g(u) &= -\frac{\chi^4(u-1)^2}{(\chi-1)^2(\chi^2 + 1)},
\quad h(u) = \frac{u^2 - 1}{\chi^2(\chi^2 - 1)},\, \text{if} \, u > 1.\\
\end{split}
\end{equation}
The section of the corresponding ellipsoid with  $\varepsilon=0.4$ is shown on the figure 
\ref{fig:ex-3}. Parameters $A$ and $q$ are rational functions of $\chi$. Their values
at $\varepsilon=0.4$ are $A\approx -0.30$ and $q\approx 0.09$.

\section{Conclusion}\label{conclusion}
Let us summarize the main results.
\begin{itemize}
\item[-] The outer gravitational potential of any level ellipsoid of revolution of 
positive mass, with the inward direction of attracting force on its surface
can be represented by the potential of the simple layer with the density
\begin{equation*}
\varrho(v, \lambda) = \frac{1+Bv^2}{\sqrt{1 - \varepsilon^2 + \varepsilon^2v^2}},
\end{equation*}
parameterized by a constant $B \geqslant -1$. The Clairaut parameter $q$ is expressed by the
function \eqref{sl-q}, depending on $B$ and the ellipsoid's eccentricity.
\item[-] The ellipsoid of revolution with the density 
\begin{equation*}
\varrho(u, v, \lambda) = 
f + g\frac{1+h v^2 }{u^2 + v^2},
\end{equation*}
where $f, g, h$ are Riemann integrable functions of $u$ is a level ellipsoid for some $q$
which depends on parameter functions and the eccentricity. 
\item[-] For the level ellipsoid with density parameters $f=0$, $g = 1$, $h=h_0 > 0$,
the Clairaut parameter $q$  is separated from zero for an arbitrary small eccentricity.
However, this example is quite artificial: the equidensites of the figure are not closed,
and there is an essential singularity of the density on the focal circle.
\end{itemize}

\section{Acknowledgements}\label{acknowledgements}
The author will be forever grateful to Professor K.~V. Kholshevnikov (1939 --- 2021), who pointed
him the direction of this research and gave a lot of invaluable advices.

\section{Appendix}\label{appendix}
The function $H(\varepsilon)$ defined by \eqref{sl-H} decreases at $[0, 1]$. 
Let us express the dependence of $H$ from $\varepsilon$ explicitly, and make sure that its derivative
is negative.
$$
H(\varepsilon) = \frac{1}{\varepsilon^5}
\left(3(2\varepsilon^4 - 15\varepsilon^2 + 9)\sqrt{1 - \varepsilon^2}\arcsin\varepsilon - \right.
\left.8\varepsilon^5 + 18\varepsilon^3 - 9\varepsilon\right)
$$
\begin{equation*}
\begin{split}
H'(\varepsilon) &= \frac{3}{\varepsilon^6}
\left(3(4\varepsilon^2 - 5)\sqrt{1 - \varepsilon^2}\arcsin\varepsilon + 2\varepsilon^5 - 17\varepsilon^3 + 15\varepsilon\right) \\&= 
\frac{3\sqrt{1-\varepsilon^2}}{\varepsilon^7}
\left((15 - 2\varepsilon^2)\sqrt{1-\varepsilon^2} - 3(5 - 4\varepsilon^2)\frac{\arcsin\varepsilon}{\varepsilon}\right)
\end{split}
\end{equation*}
Let us expand the root and the arcsine into series up to the fourth degree of $\varepsilon^2$:
\begin{equation*}
\begin{split}
(&15 - 2\varepsilon^2)\sqrt{1-\varepsilon^2} -
3(5 - 4\varepsilon^2)\frac{\arcsin\varepsilon}{\varepsilon} <\\
(&15 - 2\varepsilon^2)\left(1 - \frac{\varepsilon^2}{2} - \frac{\varepsilon^4}{8} - \frac{\varepsilon^6}{16} - \frac{5\varepsilon^8}{128}\right) -  
3(5 - 4\varepsilon^2)\left(1 + \frac{\varepsilon^2}{6} + \frac{3\varepsilon^4}{40} + \frac{5\varepsilon^6}{112} + \frac{35\varepsilon^8}{1152}\right) \\
&=\varepsilon^6\left(\frac{85}{192}\varepsilon^2 - \frac{8}{21}\varepsilon - \frac{16}{35}\right)
\end{split}
\end{equation*}
The maximum of the square trinomial in the last expression is attained at $\varepsilon=1$ 
and equals to $-2657/6720$.



\begin{thebibliography}{7}

\bibitem[Antonov 1988]{Antonov-1988}
Antonov, V.,A. and Timoshkova, E.,I. \& Kholshevnikov, K.,V.\\
Introduction to the Theory of Newtonian Potential (in Russian)\\
Moscow, Nauka, 1988

\bibitem[Hofmann-Wellenhof 2006]{Hofmann-Wellenhof-2006}
Hofmann-Wellenhof, B., \& Moritz, H. \\
Physical geodesy\\
Springer Science \& Business Media, 2006

\bibitem[Kholshevnikov 2017]{Kholshevnikov-2017}
Kholshevnikov, K.V., Milanov, D.V. \& Shaidulin, V.S. \\
The Laplace series of ellipsoidal figures of revolution \\
Vestnik St.Petersb. Univ.Math. 50, 406-413 (2017). https://doi.org/10.3103/S1063454117040112

\bibitem[Kholshevnikov 2018]{Kholshevnikov-2018}
Kholshevnikov, K.V., Milanov, D.V. \& Shaidulin, V.S. \\
Laplace series for the level ellipsoid of revolution \\
Celest Mech Dyn Astr 130, 64 (2018). https://doi.org/10.1007/s10569-018-9851-7

\bibitem[Kondratiev 2003]{Kondratiev-2003}
B. P. Kondratiev\\
Theory of Potential, and Figures of Equilibrium \\
Inst. Kosm. Res., Moscow, 2003 [in Russian]

\bibitem[Korn 1968]{Korn-1968}
Korn, G. A., \& Korn, T. M.\\
Mathematical handbook for scientists and engineers: definitions, theorems, and formulas for reference and review. \\
McGraw-Hill, 1968

\bibitem[Pizzetti 1913]{Pizzetti-1913}
Pizzetti P. \\
Principii della teoria meccanica della figura dei pianeti. \\
Pisa: E. Spoerri, 1913


\end{thebibliography}

\end{document}